\documentclass[%
 aip,
cp,  
 amsmath,amssymb,nobibnotes,
 reprint,%
]{revtex4-2}

\usepackage{graphicx}
\usepackage{dcolumn}
\usepackage{bm}

\usepackage[utf8]{inputenc}
\usepackage[T1]{fontenc}
\usepackage{mathptmx} 
\usepackage{mathrsfs}  
\usepackage{upgreek}

\def\d{{\rm d}}
\def\ds{{{\rm ds}^2}}

\newcounter{manualfn}

\begin{document}

\title{Nonperfect Carrollian Fluids Through Holography}

\author{Felipe Diaz} 
 \email{fdiaz@itmp.msu.ru}
\affiliation{
  Institute for Theoretical and Mathematical Physics, Lomonosov Moscow State University, 119991 Moscow, Russia
}

\date{\today} 

\begin{abstract}
We embed the covariant, gauge-invariant gravitational radiation criteria of Fern\'andez-\'Alvarez and Senovilla, based on conformal geometry and the Bel-Robinson tensor, into the hydrodynamic framework of gauge/gravity duality. This construction uncovers a direct correspondence between bulk gravitational waves and dissipative processes in the boundary fluid, from which a natural notion of entropy production emerges. We further analyze a smooth flat-space limit in which the dual fluid becomes Carrollian, with dissipation governed by Carroll-covariant tensors. As an example, we apply our framework to the Robinson–Trautman family of solutions.

\end{abstract}

\maketitle

\section{\label{sec:level1}Introduction}

The systematic construction of physically meaningful charges associated with gauge symmetries can be achieved through the analysis of \emph{asymptotic symmetries}. Conserved charges play a central role in probing the dynamics of any theory. A general framework to identify symmetries and their associated conserved quantities in generic spacetimes consists of specifying appropriate boundary conditions in an asymptotic region. In Yang–Mills theories, these charges reproduce electric and magnetic charges, while in general relativity they provide well-defined notions of energy and momentum \cite{Barnich:2001jy}, thereby yielding physically relevant observables. Such charges are formulated as surface integrals evaluated on the boundaries of spacetime and depend crucially on the chosen boundary conditions. Consequently, a physical theory is characterized not only by its action and bulk equations of motion, but also by the admissible boundary conditions, which determine the spectrum of observables, thermodynamic properties, and stability \cite{Barnich:2007bf}. Boundary conditions therefore serve as the interface between the mathematical structure of gravity and physical reality, enabling the meaningful study of phenomena ranging from black holes to cosmology and quantum gravity.

For three-dimensional anti-de Sitter (AdS) gravity, Brown and Henneaux \cite{Brown:1986nw} showed that under a class of boundary conditions that includes the Ba{\~n}ados-Teitelboim-Zanelli (BTZ) black hole \cite{Banados:1992wn}, the asymptotic symmetry algebra consists of two copies of the Virasoro algebra with non-trivial central extensions. These central charges famously allow for a microscopic derivation of the BTZ black hole entropy \cite{Strominger:1997eq}. This structure imposes strong constraints on the possible dual descriptions and supports the conjecture that a two-dimensional Conformal Field Theory (CFT) resides on the boundary of AdS. 

More broadly, the AdS/CFT correspondence \cite{Maldacena:1997re,Witten:1998qj,Gubser:1998bc} provides a non-perturbative framework for defining quantum gravity. Extending this paradigm to asymptotically locally de Sitter spacetimes or asymptotically flat spacetimes through a controlled limiting procedure offers a promising route toward understanding the quantum nature of the universe we inhabit.

In asymptotically flat spacetimes, the study of asymptotic symmetries was originally motivated by the attempt to understand gravitational radiation far from the strong-gravity regime \cite{Bondi:1962px, Sachs:1961zz, Blanchet:1985sp}. A key result is that the asymptotic symmetry group is not the Poincar\'e group but an infinite-dimensional extension known as the Bondi-van der Burg-Metzner-Sachs (BMS) group at null conformal infinity. Interest in the BMS group has been revived by the striking observation that its generators commute with the graviton S-matrix \cite{Strominger:2017zoo}.

Notably, one can consider a flat limit of the AdS$_3$/CFT$_2$ correspondence by performing a contraction, in the  \.{I}n\"on\"u-Wigner sense, of the boundary Virasoro generators. This procedure yields the $\mathfrak{bms}_3$ algebra \cite{Barnich:2012aw}, which underlies the lower-dimensional BMS/CFT correspondence \cite{Barnich:2010eb,Barnich:2011mi} and makes explicit the connection between asymptotic observables in AdS and those in asymptotically flat spacetimes. However, obtaining a non-singular limit at the level of bulk fields requires a suitable null gauge: the Fefferman-Graham gauge \cite{Fefferman:1984asd}, which is natural in the context of gauge/gravity duality, becomes singular in the flat limit.

A similar geometric analysis can be carried out in four dimensions. Taking a flat limit through suitable gauge choices allows one to relate the boundary of AdS$_4$ to the conformal boundary of Minkowski space. Because this hypersurface is null, its description requires not only its degenerate metric but also a vector field spanning the kernel of that metric. The resulting structure is a Carroll geometry (see \cite{Ciambelli:2025unn} for a clear review). The Carroll group was originally introduced by L\'evy-Leblond \cite{Levy-Leblond:1965dsc} and Sen Gupta \cite{SenGupta:1966qer} as the contraction of the Poincar\'e group corresponding to the vanishing speed-of-light limit, and it naturally appears in the study of physics on null hypersurfaces. Its conformal extension is, in fact, isomorphic to the BMS group \cite{Duval:2014uva}.

In particular, the flat limit of AdS/CFT in the bulk corresponds to an ultrarelativistic limit in the dual CFT, giving rise to what is now known as a Carrollian CFT; see, e.g., \cite{Bagchi:2009my,Bagchi:2016bcd,Ciambelli:2018wre,Poole:2018koa,Hijano:2019qmi,Compere:2019bua,Compere:2020lrt,Hijano:2020szl,Komatsu:2020sag,Li:2021snj,Ponomarev:2022qkx,Duary:2022pyv,Campoleoni:2023fug,Bagchi:2023fbj,Alday:2024yyj,Kraus:2024gso,Fareghbal:2024lqa,Duary:2024kxl,Marotta:2024sce,Lipstein:2025jfj,Surubaru:2025fmg,Hao:2025btl}. For reviews on flat-space holography and Carrollian CFTs, see \cite{Nguyen:2025zhg,Bagchi:2025vri}. 

A key ingredient of Carrollian holography is the inclusion of radiation \cite{Donnay:2021wrk,Donnay:2022aba,Donnay:2022wvx}. Defining radiative spacetimes with AdS asymptotics and incorporating them into a holographic framework could therefore provide a promising route to extend the lessons of AdS/CFT to Ricci-flat universes. However, defining radiative geometries in AdS is notoriously difficult due to the reflective nature of the AdS boundary and the requirement that the boundary metric remain nondynamical in order to define a consistent dual field theory (see \cite{Compere:2008us,deHaro:2008gp,Miskovic:2009bm,McNees:2024iyu,Sheikh-Jabbari:2025kjd,Parvizi:2025shq,Parvizi:2025wsg,McNees:2025acf} for discussions of these issues). 

An elegant criterion for determining whether gravitational radiation reaches infinity in the presence of a cosmological constant was developed by Fern\'andez-\'Alvarez and Senovilla (FS) in a series of works \cite{Fernandez-Alvarez:2019kdd,Fernandez-Alvarez:2020hsv,Fernandez-Alvarez:2024bkf,Fernandez-Alvarez:2025qqx}. Their construction, rooted in conformal geometry and the Bel-Robinson tensor \cite{Bel1, Robinson:1997}, provides a covariant framework for characterizing radiation in spacetimes with and without a cosmological constant. 

We embed the FS criterion into the hydrodynamic limit of the AdS/CFT correspondence \cite{Hubeny:2011hd} and analyze the interplay between bulk radiation and boundary dissipation. This yields striking holographic relations connecting gravitational waves in the bulk to out-of-equilibrium conformal dynamics at the boundary. Finally, by taking the flat limit of these holographic quantities, we obtain a new Carroll-covariant description of gravitational radiation.

This contribution is a proceedings article based on the results presented in Ref. \cite{Arenas-Henriquez:2025rpt}, where the detailed derivations were originally carried out. Here we summarize the main ideas and emphasize their interpretation within Carrollian holography and gravitational radiation.|

\section{Gravitational Radiation with Negative Cosmological Constant and Boundary Hydrodynamics}\label{Sec:FS}

Because the conformal boundary of AdS is timelike, defining outgoing gravitational radiation in asymptotically locally AdS (AlAdS) spacetimes is notoriously subtle. The AdS/CFT correspondence further enforces Dirichlet boundary conditions on the boundary metric’s conformal class, rendering the boundary reflective and preventing energy leakage to timelike infinity. Even so, the Penrose-Rindler prescription \cite{Penrose:1986ca} can still be employed: one studies the parallel transport of a tetrad along null geodesics reaching the boundary and checks that it becomes misaligned with ingoing boundary waves (see \cite{Podolsky:2003gm} for an example). In a series of papers \cite{Fernandez-Alvarez:2019kdd, Fernandez-Alvarez:2020hsv, Fernandez-Alvarez:2024bkf, Fernandez-Alvarez:2025qqx}, FS have proposed a different gauge-invariant approach to understand  gravitational radiation with negative cosmological constant, based on conformal geometry. A key object of the method is the Bel-Robinson tensor \cite{Bel1, Robinson:1997}
\begin{align}
    {\cal D}_{\mu\nu\rho\sigma} = {W}^\beta{}_{\rho\mu\alpha}{W}^\alpha{}_{\nu\sigma\beta} + \ast{W}^\beta{}_{\rho\mu\alpha}\ast\!{W}^\alpha{}_{\nu\sigma\beta}\,,
\end{align}
with $W^\mu{}_{\nu\rho\lambda}$ the Weyl tensor and the star denoting Hodge dualization, which can be seen as the gravitational analogue of the electromagnetic stress tensor as it is covariantly conserved, has the correct symmetries, and measures tidal forces at the boundary. Then, due to the fast decay of the Bel-Robinson tensor, FS conformally compactify the spacetime and compute the tensor on the unphysical spacetime $\hat{g}=\Omega^2 g$. Then, following the example of electromagnetism, one can construct the analogue of the Poynting vector that measures the flux of tidal energy at infinity. The criterion is equivalent to the presence of a Bondi news tensor in the case of $\Lambda = 0$ and can be straightforwardly extended to the case of $\Lambda > 0$~, where one finds that the electric and magnetic part of the Weyl tensor must be simultaneously diagonalizable. In the case of negative cosmological constant, which we parametrize by $\Lambda = -3\upkappa^2$~, the boundary is a Lorentzian manifold,; therefore, there is no unique non-spacelike vector at the boundary. Nonetheless, one can use any $\hat{v}$ satisfying $\hat{n}(\hat{v}) = -1$~, with $\hat{n}^\mu$ the unit normal to the boundary, such that the gravitational radiation criteria is given by 
\begin{align}\label{FScrit}
    \hat{P}({\hat{n}}) = 0~,
\end{align}
with 
\begin{align}\label{Pmu}
    \hat{P}^\mu = -\Omega^2\hat{D}^\mu{}_{\alpha\beta\gamma}\hat{v}^\alpha\hat{v}^\beta\hat{v}^\gamma~,
\end{align}
which is covariantly conserved on shell at the boundary for vacuum solutions \cite{Fernandez-Alvarez:2025qqx}.

The FS criteria for AlAdS solutions, given by \eqref{FScrit}, now translates into two conditions for the absence of radiation. A spacetime lacks radiative properties if the boundary Cotton-York tensor\footnote{Here $\epsilon_{ijk}$ is the Levi-Civita tensor, $R_{ij}^{(0)}$ and $R^{(0)}$ are the Ricci curvatures of the boundary metric $g_{(0)}$~, and $\nabla^{(0)}$ is its covariant derivative with respect to the Levi-Civita connection.}
\begin{align}
   \mathscr{C}_{ij} = \epsilon_i{}^{kl}\nabla^{(0)}_k\left(R_{jl}^{(0)} - \tfrac14R^{(0)}g^{(0)}_{jl}\right)~,
\end{align}
and the holographic stress tensor $T_{ij}$ span the same one-dimensional subspace of tensors at the boundary. Then, both are pointwise linearly dependent, i.e.
\begin{align}
    \boldsymbol{\alpha}(x^k)\mathscr{C}_{ij} = \boldsymbol{\beta}(x^k)T_{ij}~,
\end{align}
with $\boldsymbol{\alpha}$ and $\boldsymbol{\beta}$ two finite, simultaneously non-vanishing functions of the boundary coordinates. Secondly, if the contracted version of \eqref{FScrit}, that defines the \textit{radiative vector} \cite{Ciambelli:2024kre}
\begin{equation}\label{Prad}
    \tfrac{1}{16\pi G}\hat{\mathscr{P}}^{i}=\upkappa^2\mathscr{C}_{j}^{\hphantom{j}k}T_{kl}\varepsilon^{jli}~,
\end{equation}
must vanish at the boundary. 
If either of these two conditions is not satisfied, the spacetime radiates.

The radiative vector also appears by replacing $\hat{v}$ with $\hat{n}$ in \eqref{Pmu} and taking its boundary value. This quantity was previously introduced in \cite{Ciambelli:2024kre}, but its trivial value is not sufficient to determine whether radiation is present. For a spacetime to be nonradiative, both conditions must be satisfied simultaneously. For instance, in \cite{Fernandez-Alvarez:2025qqx} an example is constructed in which the radiative vector vanishes while the boundary tensors remain linearly independent, indicating that gravitational radiation is still present.

In contrast to the Ricci-flat case, the presence of radiation forces the conformal boundary of AlAdS spacetimes to be nonconformally flat. For asymptotically flat geometries, the asymptotic symmetry group is enlarged to the infinite-dimensional BMS group, which includes supertranslations. These relate the gravitational vacua before and after the passage of a wave, as the radiation shifts the asymptotic shear by an amount proportional to the gravitational memory (see \cite{Strominger:2017zoo} for a review and references therein).

In what follows, we will use the framework developed in \cite{Ciambelli:2017wou, Ciambelli:2018wre, Ciambelli:2019lap,Campoleoni:2023fug} to take the flat limit of the fluid/gravity correspondence \cite{Hubeny:2011hd} and extract information about the putative sourced Carrollian holographic field theories living on the null boundary of Ricci-flat spacetimes from the flat limit of the hydrodynamic regime of the AdS/CFT correspondence. 

\subsection{Entropy Production}

The timelike congruence $u^i$ can be interpreted as the dual fluid velocity, whose irreducible decomposition provides the kinematical properties of the fluid (see \cite{Arenas-Henriquez:2025rpt} for details). The holographic stress tensor can then be expanded as
\begin{align}
\upkappa^2T_{ij}=(\varepsilon+p)u_iu_j+ pg^{(0)}_{ij}+\tau_{ij}+u_iq_j+u_jq_i\,,
\end{align}
where $\varepsilon$ represents the local energy density, $p$ the local pressure, $q_i$ the heat current, and $\tau_{ij}$ the viscous stress tensor. Similarly, we can decompose the Cotton-York tensor along the timelike congruence in a similar way \cite{Campoleoni:2023fug}, i.e.
\begin{align}
\upkappa\mathscr{C}_{ij}=\tfrac{c}{2}\left(g^{(0)}_{ij}+\tfrac{3}{\upkappa^2} u_iu_j\right)- c_{ij}+u_ic_j+ u_jc_i\,,
\end{align}
where the descendants $c$, $c_i$, and $c_{ij}$ are referred to as the Cotton density, Cotton current, and Cotton stress tensor, respectively.

Using this expansion, the radiative vector for algebraically special Petrov-type spacetimes becomes \cite{Arenas-Henriquez:2025rpt}
\begin{equation}\label{P_aesp}
\tfrac{1}{8\pi G}\hat{\mathscr{P}}^i
=
-3\upkappa^2\left(8\pi G\varepsilon q^i-c\ast\!q^i\right)
+
16\pi G\upkappa^2\left(2\tau^{ij}q_j+u^i\left(\tau^{kl}\tau_{kl}-\tfrac{1}{\upkappa^2}q^kq_k\right)\right)\,.
\end{equation}
This expression is nontrivial only for non-perfect fluids.

Finally, using the covariant conservation of $\hat{F}^\mu=-\Omega^2\hat{\cal D}^\mu{}_{\alpha\beta\gamma}\hat{n}^\alpha\hat{n}^\beta\hat{n}^\gamma$ at the boundary \cite{Arenas-Henriquez:2025rpt}, the conservation law can be rewritten in terms of the radiative vector, yielding
\begin{align}\label{FluxLaw}
\nabla^{(0)}_i\left(\hat{\mathscr{P}}^i+Y^{(0)}u^i\right)= \hat{F}^r_{(1)}\,,
\end{align}
with
\begin{align}
Y_{(0)} = -64\pi^2 G^2\upkappa^2T^{ij}T_{ij} - \mathscr{C}^{ij}\mathscr{C}_{ij}\,, \qquad 
\hat{F}^{r}_{(1)} = -16\pi G\upkappa^2T^{i}{}_j{\mathscr{E}}^j_{(1)}{}_i - 2\upkappa\mathscr{C}^i{}_j \mathscr{M}^j_{(1)}{}_{i}\,,
\end{align}
where we have used the asymptotic expansion of the electric and magnetic parts of the Weyl tensor\footnote{To be precise, the electric and magnetic parts of the Weyl tensor are defined by decomposing the Weyl curvature with respect to a timelike observer. In our case, since the boundary metric is Lorentzian, the decomposition is performed with respect to a spacelike vector instead. Nevertheless, within the AdS/CFT framework it remains standard practice to refer to these components as the ``electric'' and ``magnetic'' parts of the Weyl tensor.}, i.e.
\begin{align}
{\mathscr{E}}^\mu{}_{\nu} :={}&-\hat{W}^\mu{}_{\alpha\beta\nu}\hat{n}^\alpha \hat{n}^\beta
= r^{-1}{\mathscr{E}}^\mu_{(0)}{}_\nu + r^{-2}{\mathscr{E}}^\mu_{(1)}{}_\nu + \dots\,,\\
{\mathscr{M}}^\mu{}_{\nu} :={}&-\ast \hat{W}^\mu{}_{\alpha\beta\nu}\hat{n}^\alpha \hat{n}^\beta
= r^{-1}{\mathscr{M}}^\mu_{(0)}{}_\nu + r^{-2}{\mathscr{M}}^\mu_{(1)}{}_\nu + \dots\,,
\end{align}

Then the flux law \eqref{FluxLaw}, written in terms of fluid variables, gives the entropy production criterion \cite{Arenas-Henriquez:2025rpt}
\begin{align}\label{SprodEq}
\nabla_i^{(0)}(\varepsilon T s^i)
=
\nabla^{(0)}_i\!\left(\tfrac43 \tau^{ij}q_j + \tfrac{2}{3\upkappa^2}q^j q_j u^i + \varepsilon^2 u^i - \tfrac{c^2}{16\pi G} u^i - \tfrac{c}{8\pi G}\ast q^i\right)
- \tfrac{1}{192\pi^2 G^2 \upkappa^2}\hat{F}^r_{(1)}\,,
\end{align}
where $s^{i} = \tfrac{1}{T}\!\left(\tfrac{3}{2}\varepsilon u^{i} + q^{i}\right)$ is the boundary first-order entropy current \cite{Kovtun:2012rj}. This relation highlights the connection between out-of-equilibrium conformal dynamics at the boundary and bulk gravitational radiation encoded in the radiative vector.

\section{Carrollian Limits}

In order to obtain a Ricci-flat spacetime from a limiting process, let us consider the Papapetrou-Randers parametrization of the boundary metric
\begin{align}
\ds = -{\upkappa^2}\left(\Omega\d u - b_A \d x^A\right)^2 + a_{AB}\d x^A \d x^B\,,
\end{align}
which is useful to describe holographic Carrollian fluids in the flat limit \cite{Ciambelli:2018wre,Ciambelli:2019lap,Mittal:2022ywl,Petkou:2022bmz,Campoleoni:2023fug,Miskovic:2023zfz}, as it is stable under Carrollian diffeomorphisms and becomes degenerate in the flat limit. Here $a_{AB}$ and $\Omega$ transform, respectively, as a Carroll rank-2 tensor and a Carroll scalar, while $b_A$ transforms inhomogeneously. 

In the ultralocal limit $\upkappa\to0$, the resulting null geometry is endowed with a degenerate metric whose kernel is generated by the Carroll vector field $\upnu$, whose fibre structure is aligned with the time coordinate $u$, with a dual 1-form $\upmu$ known as the clock form such that $\upnu(\upmu)=-1$, and $b_A$ serves as an Ehresmann connection \cite{Ciambelli:2019lap}. We can equip the Carroll structure with an ambiguous connection that parallel transports the degenerate metric and the Carroll vector to form a strong Carroll structure \cite{Duval:2014uva} (see \cite{Ciambelli:2025unn} for a review on Carrollian manifolds). 

We follow this approach and use such a strong structure together with a time/space splitting formalism (see for instance \cite{Ciambelli:2017wou,Ciambelli:2018wre,Ciambelli:2019lap}), where one chooses the connection to construct a pair of covariant derivatives $\{\tfrac{1}{\Omega}\tilde{\cal D}_u,\tilde{\cal D}_A\}$ along the base and vertical components of the Carrollian manifold such that acting on tensors maintains their transformation properties under Carrollian diffeomorphisms and Weyl rescalings. 

The flat limit of the fluid/gravity correspondence in this framework has been extensively studied in \cite{Campoleoni:2023fug}. We now apply the same formalism to the radiative vector \eqref{Prad} to relate gravitational radiation in Ricci-flat spacetimes to Carrollian hydrodynamics at the null boundary. 

Expanding in powers of $\upkappa$ the boundary Cotton tensor and stress tensor through the decomposition of \cite{Campoleoni:2023fug}, we find that the nonrelativistic limit of the radiative vector gives rise to two nontrivial Carroll-covariant finite contributions
\begin{align}
\hat{\varrho} = \lim_{\upkappa\to0} \tfrac{1}{\Omega}\hat{\cal P}_u = -128\pi^2 G^2 \Sigma_{AB}\Sigma^{AB}\,,\qquad 
\hat\Upsilon^A = \lim_{\upkappa\to 0}\hat{\cal P}^A = -256\pi^2 G^2 \Sigma^{AB}Q_B\,,
\end{align}
where transverse indices are raised and lowered with $a_{AB}$, and 
\begin{align}
Q^A = \lim_{\upkappa\to0}q^A\,,\qquad 
\Sigma_{AB} = -\lim_{\upkappa\to 0}\upkappa^2\tau_{AB}\,,
\end{align}
are the Carroll energy flux and Carroll viscous stress tensor, respectively.

The pair $\{\hat{\varrho},\hat\Upsilon^A\}$ forms a Carroll pair that we refer to as the Carroll radiative scalar and Carroll radiative vector, respectively. Both depend on the Carrollian viscous stress tensor $\Sigma_{AB}$ and heat current $Q^A$, indicating that bulk radiation generates dissipation in the boundary theory, with the Carroll boundary data $(Q^A,\Sigma^{AB})$ carrying the information of bulk radiation. Another example of sources of boundary dissipation are angular momentum and NUT charges. Then, the Kerr-NUT black hole would be dual to a nonperfect Carrollian fluid. Nonetheless, the solution does not have a Carrollian viscous stress tensor $\Sigma_{AB}$ \cite{Arenas-Henriquez:2025rpt}, which is consistent with the fact that such a spacetime does not radiate.

Furthermore, comparing the flat limit of the radiative vector obtained through the NU gauge with the one obtained in Bondi coordinates \cite{Fernandez-Alvarez:2019kdd}, it is suggested to identify the dissipative corrections with the Bondi news tensor as
\begin{align}
\Sigma_{AB} \sim \dot{N}_{AB}\,,\qquad 
Q^A \sim D_B^{(0)}N^{AB}\,,
\end{align}
where the dot indicates a derivative with respect to the Bondi time and $D_A^{(0)}$ is the covariant derivative with respect to the transverse metric of null infinity in Bondi coordinates.

\subsection{Example: Robinson-Trautman Spacetimes}

Let us now turn to the Robinson-Trautman family of exact solutions, defined by the presence of an expanding, twist-free, and shear-free null geodesic congruence. These geometries belong to the algebraically special sector of the Petrov classification and describe spherical gravitational radiation propagating along a preferred null direction. In Newman-Unti coordinates, the line element reads
\begin{align}\label{dsRT}
    \ds = -\left(- \tfrac{\Lambda}{3}r^2+ \dot{\Phi}(u,x^A)r-\Delta\Phi(u,x^A) - \tfrac{2m}{r}\right)\d u^2-2\d u\d r + 2r^2e^{\Phi(u,x^A)}\d \zeta \d \bar{\zeta}\,, 
\end{align}
where the dot denotes a derivative with respect to the retarded time $u$ and $r$ is the null radial coordinate. Surfaces at constant $u$ and $r$ represent a deformed two-sphere coordinatized by conformally flat K\"ahler coordinates $x^A = (\zeta,\bar\zeta)$, with the associated Laplacian given by $\Delta = e^{-\Phi}\partial\bar\partial$, where $\partial \equiv \partial_\zeta$ and $\bar\partial \equiv \partial_{\bar{\zeta}}$. The solution is asymptotically (locally) flat, dS, or AdS according to whether the cosmological constant is zero, positive, or negative, respectively. Here $m$ is an integration constant related to the physical energy of the configuration\footnote{In principle, the mass parameter $m$ can depend on $u$. However, the line element is stable under a class of suitable coordinate transformations that allow us to reach a gauge that makes $m$ constant.}. Einstein's equations impose that the Robinson-Trautman (RT) field $\Phi$ satisfies the fourth-order parabolic equation\footnote{The higher-derivative character of the RT equation arises because the metric contains an explicit dependence on the Laplacian of the deformed two-sphere.}
\begin{align}
    \Delta\Delta\Phi + 3m\dot{\Phi} = 0\,,
\end{align}
which is independent of $\Lambda$ and is known as the RT equation and coincides with the Calabi flow on the two-sphere. 

Let us now consider the case $\Lambda  = -3\upkappa^{2} <0$ and use the FS criteria to check which subspace corresponds to radiative spacetimes and to extract holographic information from them. 

The only kinematical quantity is the expansion given by $\Theta_{\rm RT} = \dot\Phi$. The boundary geometry is given by
\begin{align}
    g_{(0)}(\d x,\d x) = - {\upkappa^2}\d u^2+2e^\Phi\d \zeta\d\bar\zeta\,,
\end{align}
which is nonconformally flat since its Cotton tensor does not vanish in general and becomes degenerate in the $\upkappa\to0$ limit. The Papapetrou-Randers coefficients are given by
\begin{align}
    \Omega = 1\,,\qquad b_A = \vec{0}\,,\qquad a_{AB} = 2e^\Phi\delta^\zeta_A\delta^{\bar\zeta}_B\,,
\end{align}
such that the boundary timelike velocity 
\begin{align}
    u^i\partial_i = \partial_u\,,\qquad u_i\d  x^i = -{\upkappa^2}\d u\,,
\end{align}
with norm $g_{(0)}(u,u)=-\upkappa^{2}$, defines the decomposition for the boundary Cotton-York and stress tensor \cite{Campoleoni:2023fug} with descendants 
\begin{align}
c ={}&0\,,  &&\varepsilon ={} \tfrac{m}{4\pi G}\,, \\
    c_u ={}&0\,, && q^u ={} 0\,,\\ 
c_\zeta={}&\tfrac{i}{2}\left(\partial\Phi\Delta\Phi-\Delta\partial\Phi\right)\,, && q^\zeta = \tfrac{e^{-\Phi}}{16\pi G}\left(\Delta\bar\partial\Phi-\bar\partial\Phi\Delta\Phi\right)\,,\\ 
c_{\bar\zeta} ={}& \tfrac{i}{2}\left(\Delta\bar\partial\Phi-\bar\partial\Phi\Delta\Phi \right)\,, && q^{\bar\zeta} ={} \tfrac{e^{-\Phi}}{16\pi G}\left(\Delta\partial\Phi-\partial\Phi\Delta\Phi\right)\,,\\ 
c_{ij} ={}& \tfrac{i}{2}{\rm diag}\left[0,\partial\Phi\partial\dot\Phi-\partial^2\Phi,\bar\partial\Phi\bar\partial\dot\Phi-\bar\partial^2\Phi\right]\,, && 
\tau_{ij}={} \tfrac{1}{16\pi G \upkappa^2}{\rm diag}\left[0,\partial\Phi\partial\bar\partial\Phi-\partial^2\dot\Phi,\bar\partial\Phi\bar\partial\dot\Phi-\bar\partial\dot\Phi\right]\,,
\end{align}
and the components of the boundary radiative vector are
\begin{align}
\hat{\mathscr{P}}^u ={}& e^{-2\Phi}\left[{\upkappa^{-2}} \left(\bar\partial\Phi\bar\partial\dot\Phi - \bar\partial^2\dot\Phi\right)\left(\partial\Phi\partial\dot\Phi - \partial^2\dot\Phi\right)- \left(\bar\partial\Phi\Delta\Phi - \Delta\bar\partial\Phi\right)\left(\partial\Phi\Delta\Phi - \Delta\partial\Phi\right)\right]\,, \\
\hat{\mathscr{P}}^{\zeta}={}& e^{-2\Phi}\left[{3m}{\upkappa^2}e^\Phi\left(\bar\partial\Phi\Delta\Phi - \Delta\bar\partial\Phi\right) - \left(\Delta\partial\Phi - \partial\Phi\Delta\Phi\right)\left(\partial\Phi\bar\partial\dot\Phi - \bar\partial^2\dot\Phi\right)\right]\,,   \\
\hat{\mathscr{P}}^{\bar\zeta}={}& e^{-2\Phi}\left[{3m}{\upkappa^2}e^\Phi\left(\partial\Phi\Delta\Phi - \Delta\partial\Phi\right) - \left(\Delta\bar\partial\Phi - \bar\partial\Phi\Delta\Phi\right)\left(\bar\partial\Phi\partial\dot\Phi - \partial^2\dot\Phi\right)\right]\,.
\end{align}

One can check that in general the boundary Cotton-York tensor is linearly independent of the holographic stress tensor. Therefore, the non-radiative sector of this family of solutions corresponds to the case in which the radiative vector is trivial. Notice that for a time-independent solution there are still non-zero components of the radiative vector. Finally, using \eqref{SprodEq} we find that there is no first-order entropy production in the boundary fluid
\begin{align}
    \vec\nabla^{(0)} \cdot \vec{s} = 0\,,
\end{align}
off shell. Notice that although radiation is present, the dual fluid still possesses a conserved entropy current. See \cite{Ciambelli:2017wou} for a discussion. 

We now proceed to study the ultralocal flat limit $\upkappa\to0$. The field of observers is given by $\upnu = \partial_u$, with dual clock form $\upmu = -\d u$, with trivial Ehresmann connection, such that the Carrollian shear, vorticity, and acceleration remain trivial. The only nontrivial coefficients of the expansion are Carroll-covariant and are given by
\begin{align}
\chi^A ={}& \tfrac{i}{2}e^{-2\Phi}\left( \bar\partial^2\partial\Phi - \bar\partial\Phi\partial\bar\partial\Phi , \partial\Phi\partial\bar\partial\Phi - \partial^2\bar\partial\Phi\right)\,, \\
X^{AB} ={}& \tfrac{i}{2}e^{-2\Phi}{\rm diag}\left[ \bar\partial^2\dot\Phi - \bar\partial\Phi\bar\partial\dot\Phi, \partial^2\dot\Phi - \partial\Phi\partial\dot\Phi\right]\,, \\
\Sigma^{AB} ={}& \tfrac{1}{16\pi G}e^{-2\Phi}{\rm diag}\left[\bar\partial\Phi\bar\partial\dot\Phi - \bar\partial^2\dot\Phi , \partial\Phi\partial\dot\Phi - \partial^2\dot\Phi \right]\,, \\
Q_A ={}& \tfrac{1}{16\pi G}e^{-2\Phi}(\partial\bar\partial^2\Phi - \bar\partial\Phi\partial\bar\partial\Phi,\bar\partial\partial^2\Phi-\partial\Phi\partial\bar\partial\Phi)\,,
\end{align}
where transverse indices are raised and lowered with the transverse Carroll metric $a_{AB}$, and Einstein's equations reduce to $\tilde{\cal D}_A \ast \chi^A = 0$.\footnote{The Hodge dual acts on Carroll vectors as $\ast V_A = \tilde{\eta}^B{}_A V_B$ where $\tilde{\eta}_{AB} = \sqrt{a}\epsilon_{AB}$ and $\epsilon_{AB}$ is the Levi-Civita symbol.} 

Finally, the Carrollian radiative pair $(\hat\varrho,\hat\Upsilon_A)$ are given by 
\begin{align}
\hat{\varrho}={}&-e^{-2\Phi}\left(\partial^2_{\zeta}\dot{\Phi}-\partial_{\zeta}\Phi\partial_{\zeta}\dot{\Phi}\right)\left(\partial^2_{\bar{\zeta}}\dot{\Phi}-\partial_{\bar{\zeta}}\Phi\partial_{\bar{\zeta}}\dot{\Phi}\right)\,, \\
\hat{\Upsilon}_{A}={}&e^{-\Phi}\left(\left(\partial^2_{\zeta}\dot{\Phi}-\partial_{\zeta}\Phi\partial_{\zeta}\dot{\Phi}\right)\left(\partial_{\bar{\zeta}}\Phi\Delta\Phi-\Delta\partial_{\bar{\zeta}}\Phi\right),\left(\partial^2_{\bar{\zeta}}\dot{\Phi}-\partial_{\bar{\zeta}}\Phi\partial_{\bar{\zeta}}\dot{\Phi}\right)\left(\partial_{\zeta}\Phi\Delta\Phi-\Delta\partial_\zeta\Phi\right)\right)\,.
\end{align}

Notice that both quantities vanish if $\dot\Phi = 0$, indicating that the solution radiates only for time-dependent RT fields with dual nonperfect Carrollian fluids. A particularly interesting subclass within the RT family of solutions is that of accelerating black holes. In \cite{Arenas-Henriquez:2025rpt} this framework is applied to such geometries, demonstrating that radiation is inevitably produced whenever the spacetime undergoes acceleration. Their holographic properties have also been explored extensively \cite{Hubeny:2009kz,Anabalon:2018ydc,Barrientos:2022bzm,Arenas-Henriquez:2022www,Arenas-Henriquez:2023hur,Cisterna:2023qhh,Tian:2023ine,Arenas-Henriquez:2024ypo,Roychowdhury:2024oih,Luo:2024cwm,Li:2025rzl}, highlighting the rich structures that acceleration brings to the correspondence.

\section{Conclusions}

In this work, we use the covariant and gauge-invariant FS program to characterize gravitational radiation by constructing a Poynting-like vector that measures the flux of tidal energy reaching the conformal boundary. Within this framework, and employing the hydrodynamic regime of the AdS/CFT correspondence, we relate bulk gravitational radiation to a specific form of dissipation and to the production of entropy in the boundary theory. We then parametrize the boundary geometry \textit{\'a la} Papapetrou-Randers, which allows us to construct Carroll-covariant quantities at the boundary in the flat limit. In particular, we construct a radiative scalar and a radiative vector that are both Carroll-covariant and become nontrivial only in the presence of bulk gravitational radiation for Ricci-flat spacetimes. We use the Robinson-Trautman family of solutions as a key example illustrating the formalism. Remarkably, the boundary entropy current remains conserved despite bulk radiation inducing boundary dissipation, indicating that the dual fluid undergoes an isoentropic Moutier's cycle \cite{Ciambelli:2017wou}.

It would be interesting to identify a radiative bulk solution whose dual explicitly produces entropy. One possibility is to explore the full Pleba\'nski-Demia\'nski family of solutions (see, for instance, \cite{Barrientos:2023dlf} and references therein), or the generalization of Robinson-Trautman spacetimes with twist, which encompasses more general non-stationary radiative geometries. Another worthwhile direction is to consider a more general definition of the entropy current that includes second- (or higher-) order terms in the fluid velocity \cite{Rezzolla:2013dea}. This would provide a broader relation between bulk radiation and boundary dissipation.

Typically, second-order transport coefficients are introduced because first-order viscous hydrodynamics leads to superluminal modes, and higher-order terms are required to restore causality (which can be systematically studied for strongly coupled plasmas through the gauge/gravity correspondence \cite{Baier:2007ix,Kleinert:2016nav}). Moreover, we have not yet analyzed the flat limit of \eqref{SprodEq}, which naturally contains second-order contributions. We expect that studying this limit will help formulate a consistent second-order entropy current for a Carrollian fluid and clarify its putative bulk dual.

Another natural extension is to consider different spacetime dimensions or higher-curvature gravity theories, which are known to enrich the AdS/CFT correspondence and arise naturally from the string-theoretic origin of the duality. Nonetheless, our framework relies crucially on the Bel-Robinson tensor, whose generalization to other dimensions or to theories beyond GR is notoriously difficult unless supersymmetry is present \cite{Deser:1999jw}. A more general, dimension-independent notion of gravitational radiation in AlAdS spacetimes would therefore be required in order to analyze such systems holographically and to understand the corresponding boundary ultralocal limit.

Finally, given that Carrollian CFTs emerge in the tensionless limit of string theory \cite{Bagchi:2013bga,Bagchi:2015nca,Bagchi:2016yyf,Bagchi:2017cte,Bagchi:2020fpr,Bagchi:2020ats,Bagchi:2026wcu} and that higher-spin theories admit well-defined Carrollian extensions \cite{Campoleoni:2017mbt,Campoleoni:2017qot,Campoleoni:2020ejn,Campoleoni:2021blr,Bekaert:2022ipg,Liu:2023jnc}, it would be interesting to examine the flat limit of the higher-spin holographic formulation developed in \cite{Vasiliev:2012vf,Diaz:2024iuz,Diaz:2024kpr}. This approach provides an equivalent unfolding framework on both sides of the duality and allows one to circumvent the Maldacena-Zhiboedov no-go theorem \cite{Maldacena:2011jn}, opening the door to interacting CFTs endowed with gauge symmetries.

\begin{acknowledgments}
It is a pleasure to thank the organizers for arranging a very
stimulating workshop. The results reported in this paper have been obtained in
collaboration with G. Arenas-Henriquez, L. Ciambelli, W. Jia, and D. Rivera-Betancour. We are grateful to Glenn Barnich, Jos\'e Barrientos, Hern\'an Gonz\'alez, Sercan H\"{u}sn\"{u}gil, Leonardo Sanhueza, Jos\'e Senovilla, Konstantinos Siampos, and Per Sundell for useful discussions.
\vspace{5mm}
\paragraph{Conflict of Interest:} \!\!The authors declare that they have no conflicts of interest.

\end{acknowledgments}

\nocite{*}
\bibliography{aipsamp}

\end{document}